\documentclass[11pt]{article}
\usepackage{spconf}
\usepackage{blindtext}
\usepackage{graphicx}
\usepackage{amsfonts}
\usepackage{color}
\usepackage{xcolor}
\usepackage{amsmath}
\usepackage{stmaryrd}
\usepackage{algorithm, caption}
\usepackage{algpseudocode}
\usepackage{hyperref}
\usepackage{amsthm} 
\usepackage{mathtools}
\usepackage{subcaption} 
\usepackage{bm}
\usepackage{booktabs}
\usepackage{colortbl} 
\usepackage{booktabs}
\usepackage{float}
\usepackage{siunitx}
\usepackage{fixltx2e}
\usepackage{multirow}
\usepackage[T1]{fontenc}
\graphicspath{{Figures/}}

\DeclarePairedDelimiter\abs{\lvert}{\rvert}%
\DeclarePairedDelimiter\norm{\lVert}{\rVert}%

\makeatletter
\let\oldabs\abs
\def\abs{\@ifstar{\oldabs}{\oldabs*}}
\let\oldnorm\norm
\def\norm{\@ifstar{\oldnorm}{\oldnorm*}}
\makeatother

\hypersetup{
	bookmarksnumbered=true, bookmarksopen=true, colorlinks,
	citecolor=black,%
	filecolor=black,%
	linkcolor=black,%
	urlcolor=black
}

\begin{document}
\ninept

\title{Ear-EEG sensitivity modelling for neural and artifact sources}
\name{Metin Yarici, Mike Thornton, Danilo P. Mandic}
\address{Department of Electrical and Electronic Engineering, Imperial College London, SW7 2AZ, UK
\\E-mails: \{metin.yarici16,  d.mandic\}@imperial.ac.uk}

\markboth{Journal of \LaTeX\ Class Files,~Vol.~6, No.~1, January~2007}%
{Shell \MakeLowercase{\textit{et al.}}: Bare Demo of IEEEtran.cls for Journals}

\maketitle

\begin{abstract}
\noindent
The ear-EEG has emerged as a promising candidate for wearable brain monitoring in real-world scenarios. While experimental studies have validated ear-EEG in multiple scenarios, the source-sensor relationship for a variety of neural sources has not been established. In addition, a detailed theoretical analysis of the ear-EEG sensitivity to sources of artifacts is still missing. Within the present study, the sensitivity of various configurations of ear-EEG is established in the presence of neural sources from a range of brain surface locations, in addition to ocular sources for the blink, vertical saccade, and horizontal saccade eye movements which produce artifacts in the EEG signal. Results conclusively support the introduction of ear-EEG into conventional EEG paradigms for monitoring neural activity that originates from within the temporal lobes, while also revealing the extent to which ear-EEG can be used for sources further away from these regions. The use of ear-EEG for sources that are located further away from the ears is supported through the analysis of the prominence of ocular artifacts in ear-EEG. The results from this study can be used to support both existing and prospective experimental ear-EEG studies and applications in the context of both neural and ocular artifact sensitivity. \\

\vspace{2mm}
\end{abstract}
\section*{Introduction}

Electroencephalography (EEG) refers to the method of brain monitoring that utilises electric neural potentials on the surface of the scalp, whereby potentials are detected by electrodes attached to the surface of the scalp and processed by accompanying hardware. Common uses of EEG are clinical, and involve the localisation and characterisation of epilepsy related seizures \cite{noachtar2009role} and the objective assessment of hearing in infants \cite{schulman1979brain}. Outside the laboratory environment, EEG has attracted attention from multiple real world applications of brain monitoring, such as brain computer interfaces (BCI), since modern EEG hardware is available in a miniaturised, portable form \cite{wolpaw2002brain,abiri2019comprehensive}.

Conventional EEG is recorded through an array (montage) of electrodes placed across the entire scalp, termed a scalp EEG montage. As a result of the wide coverage of the human scalp achievable with a scalp EEG montage, conventional EEG offers good spatial sensitivity to a variety of neuronal activity from across the brain surface. However, conventional scalp EEG is not suited for wearable applications as a result of the difficulty of integration of the hardware with everyday activity; the use of scalp EEG is cumbersome, obtrusive, time-consuming to set up, difficult to use without specialist supervision, and introduces unwanted stigma. For these reasons, alternative, miniaturised, and wearable EEG montages which address these shortcomings hold much promise.

One such candidate is ear-EEG \cite{looney2011ear}, which employs a small number of electrodes which measure EEG from the surface of the skin on the outer-ear \cite{mikkelsen2015eeg}. Importantly, ear worn devices are familiar, naturally discreet, unobtrusive, non-stigmatising, and potentially easy-to-use, thus providing a convenient base for wearable health monitoring platforms. Ear-EEG has been shown to be a reliable alternative to scalp EEG in several settings; sleep stage classification \cite{eareeg1,eareeg1b}, drowsiness onset detection \cite{eareeg8}, objective hearing threshold estimation \cite{eareeg6}, bio-metric authentication \cite{eareeg3}, epileptic waveform detection \cite{eareeg5}, brain-computer-interfaces \cite{eareeg9}, and emotion recognition \cite{eareeg4}. Additionally, the susceptibility of ear-EEG to various artifacts has also been characterised experimentally for auditory neural activity detection in the presence of head, eye, and jaw movements \cite{eareeg10}.

\begin{figure}
    \centering
    \includegraphics{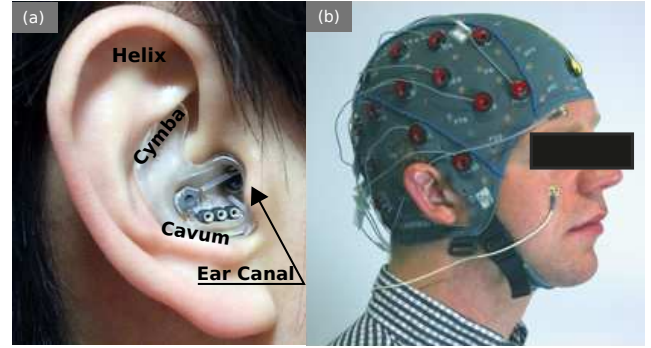}
    \caption{Images of EEG sensing technologies: (a) In-ear-EEG device mounted on the left ear of a subject. Common locations for electrodes on the ear canal, concha (cavum and cymba) and helix are shown \cite{looney2012ear}. (b) Standard scalp EEG cap and electrodes mounted on the head of a subject.}
    \label{fig eegs}
\end{figure}

However, the existing experimental literature on the ear-EEG fails to establish the source-sensor relationship for ear-EEG, i.e., the variation in sensitivity of ear-EEG corresponding to variations in neural source origins. This stems from the fact that, in practice, EEG is typically sensitive to a vast number of simultaneously active neural sources contained within some area of the brain surface. As such, when a low number of electrodes confined to a small area are used, the localisation or extraction of a single source of activity is extremely difficult \cite{grech2008review}. 

\begin{table*}[t!]
\centering
\caption{Dielectric properties of the tissue in the model. Values for conductivity ($\sigma$) and relative permittivity ($\epsilon_{r}$) are provided.}
\label{tab:tissue_props}
\resizebox{\textwidth}{!}{%
\begin{tabular}{@{}llllllll@{}}
\cmidrule(l){2-8}
                & \multicolumn{7}{c}{Tissue}                                                                        \\ \cmidrule(l){2-8} 
                & Bone  & Skin (dry) & Brain (grey matter) & Muscle & Parotid glands & Eyes (vitreous humour) & Air \\ \cmidrule(l){2-8} 
$\sigma (S/m)$ & \num{2e-2}  & \num{2e-2}    & \num{2.9e-2}             & \num{2.2e-1} & \num{6.7e-1}         & \num{1.5}                    & \num{0}  \\
$\epsilon_{r}$    & \num{5.5e-4} & \num{1.1e3}     & \num{4e7}              & \num{2.6e7} & \num{9.4e1}        & \num{9.9e1}                 & \num{1}  
\end{tabular}%
}
\end{table*}

Physics modelling of the propagation of neural potential to EEG sensors, commonly termed forward modelling, can be used to estimate the source sensor relationship for various configurations of EEG (montages/channels). In this paradigm, the electric potential on the surface of the scalp arising due to a current dipole source within the brain volume is estimated by applying Maxwell's equations to a structurally accurate dielectric model of the head \cite{sarvas1987basic}, termed a volume conductor model. High resolution imaging, such as magnetic resonance imaging (MRI) scans or computerised tomography (CT) scans enable the construction of detailed volume conductor models. Unlike experimentation, forward modelling allows for the testing of a larger number of channels and sources in a time-efficient way. Additionally, neural sources may be modelled individually, without the discussed interference from neighbouring neural sources. 

Some limited forward modelling work related to ear-EEG has been reported in the literature. In an ear-EEG modelling study \cite{Kappel2019}, Kappel \textit{et al.} demonstrated the feasibility of source localisation using in-ear-EEG and provided highly detailed subject specific forward models created by scanning each subject's head anatomy. However, the source-sensor relationship for in-ear-EEG was not evaluated in detail. In a more detailed analysis, Meiser \textit{et al.} employed forward modelling in order to compare the sensitivity of scalp EEG with CEEG - an alternative ear-EEG method which utilises the skin surface surrounding the ear \cite{Meiser2020}. The study presented in this paper aimed to establish a detailed source sensor relationship for various configurations of ear-EEG.  

While the sensitivity to neural sources is a key determining factor in the reliability of an EEG technology, equally important is its robustness in the presence of artifacts. Despite the fact that the physics mechanisms underlying most common EEG artifacts have been long established (motion \cite{symeonidou2018effects,oliveira2016induction}, eye movement \cite{gratton1998dealing,roy2014eye}, muscle activation \cite{ma2012muscle,richer2019adding,muthukumaraswamy2013high}, and electrical interference \cite{webster2009medical}), to date, there is no theoretical study of such artifacts in ear-EEG. In this paper, as well as exploring the source-sensor relationship for various neural sources, we aim to provide a detailed characterisation of the source-sensor relationship for various ocular sources of artifacts, utilising equivalent current dipole data for blinks, vertical saccades, and horizontal saccades (collected by Lins \textit{et al.} \cite{lins1993ocular2}. This paper aims to highlight the utility of approaching ear-EEG equipped with theoretical knowledge of not only neural source sensitivity, but also that for sources of artifact. 

\section*{Methods} 
Modelling was conducted in COMSOL Multiphyiscs ® - a multi-physics modelling platform which enables finite element electromagnetic modelling of various types. Geometric and dielectric human tissue data from the Information Technologies in Society Foundation (IT'IS) \cite{iacono2015mida,database_itis} was used to build an accurate (dielectric) model of the human head and ears to be used within COMSOL. The following section describes the structure of the model and the implementation of physics modelling within the COMSOL software. 

\begin{figure*}[t]
    \centering
    \includegraphics{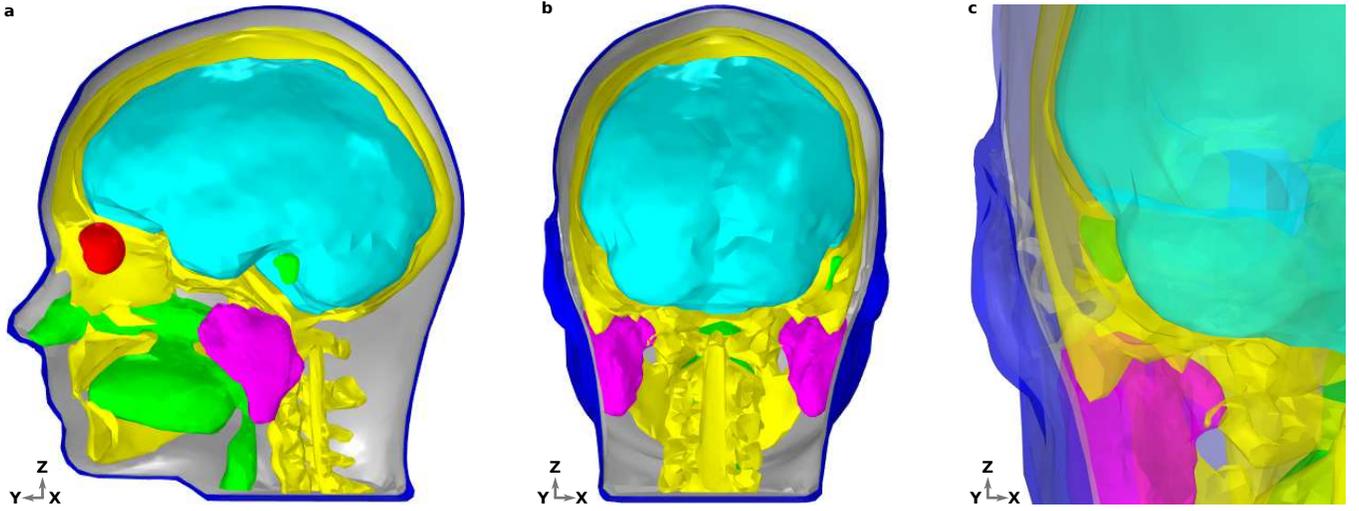}
    \caption{Dielectric distribution within the head and ear model. Different tissue groups are discernible by colour; brain-cyan, bone-yellow, muscle-grey, air-green, vitreous humour-red, skin-blue. (a) Saggital section of the head showing the interior of the model. The skin and bones are hidden to the left of the mid line sagittal plane, such that the brain, eyes, glands, and internal air are clearly visible. (b) Coronal section of the head with the skin and bone hidden to the rear of the mid line coronal plane. (c) Coronal section of the head, zoomed in to the left ear region, viewed from a posterior position. The skin and bones are hidden to the rear of the mid line coronal plane. Tissues are transparent, to enable a detailed view of the entire ear and surrounding structures}
    \label{fig model}
\end{figure*}

\subsection*{Forward modelling}
For both EEG modelling, the standard volume conductor - forward modelling approach was employed \cite{sarvas1987basic}. In addition, the same approach was employed for ocular artifact modelling. The utilisation of such an approach for ocular source modelling is motivated by the fact that the ocular sources used within this study were described as equivalent current dipoles, which were discerned through standard EEG inverse modelling procedures in the original study \cite{lins1993ocular2}. 
\subsubsection*{Volume conductor model}
 A head and ear volume conductor model was created utilising data from the MIDA head model \cite{iacono2015mida}; a highly accurate imaging data collection for an adult human head. For this study, an adapted, simplified model, including 7 different tissue types was created. The tissue types were skin, including the inner and outer dermis of the whole head, neck, and outer ear; bone, including the skull and the C1-C3 vertebrae; brain tissues including grey matter, white matter, cerebellum, and brainstem surface; internal air (respiratory tracts and mastoid air-cells within the skull); parotid glands; vitreous humour (left and right eyes); and muscle. The muscle tissue was not designated a specific geometry, rather, surrounding regions of the other tissues were endowed with muscle tissue properties. The tissue properties used within this model are provided in Table \ref{tab:tissue_props}.

The COMSOL Multiphysics software requires that geometric data does not exhibit non-manifold edges and self-intersecting faces. Therefore, the MIDA data was adapted and simplified during a pre-processing procedure in order to meet these compatibility requirements. Details of the pre-processing procedure are provided in the appendix, while the entirety of the final model geometry can be requested.

Within the COMSOL software, forward modelling was conducted within the electric currents interface (AC/DC Module) through the following methods. An equation for current density was used to solve for electric potential throughout the model:

\begin{equation}
    \mathbf{J} = \sigma\mathbf{E} + j\omega \mathbf{D} + \mathbf{J_e}, 
\end{equation}

\noindent where $\mathbf{J}$ is the total current density, $\sigma$ is material conductivity, $E$ is the electric field, $j$ is the imaginary number, $\omega$ is the frequency of current, $\mathbf{D}$ is the displacement field, and $\mathbf{J_e}$ is the external (source) current density.

The following equation of continuity was imposed across tissue boundaries:

\begin{equation}
    n_2(\mathbf{J_1} - \mathbf{J_2})=0,
\end{equation}

where $n$ is a unit vector that is normal to (and directed away from) the boundary and $\mathbf{J}$ is the electric current density. The indices $1$,$2$ describe the regions of space either side of the boundary. The head model was placed at the centre of a large (r = \SI{10}{m}) sphere with the dielectric properties of air (conductivity $\sigma =$ \SI{0}{mS \per cm}, relative permittivity $\epsilon = 1$). The surface of the sphere acted as the ground in the AC model. See \cite{com1,com2,com3} for examples of volume conductor modelling within COMSOL. 

\subsection*{Neural sources}\label{neural sig mod}
The goal of the presented modelling was to evaluate the ear-EEG source-sensor relationship for a variety of realistic neural sources. The source space was restricted to the surface of the brain and comprised 990 homogeneously distributed source locations. In this way, an exploration of the source sensor relationship for a range of realistically located neural sources was achieved. Each location was occupied by a single source, enabling a fine-grained mapping of the brain surface. Each source was orientated perpendicular to the surface of the brain. Sources were modelled as point current dipoles; the location, orientation, and magnitude of which were specified \cite{sarvas1987basic}.

 \begin{figure*}
    \centering
    \includegraphics{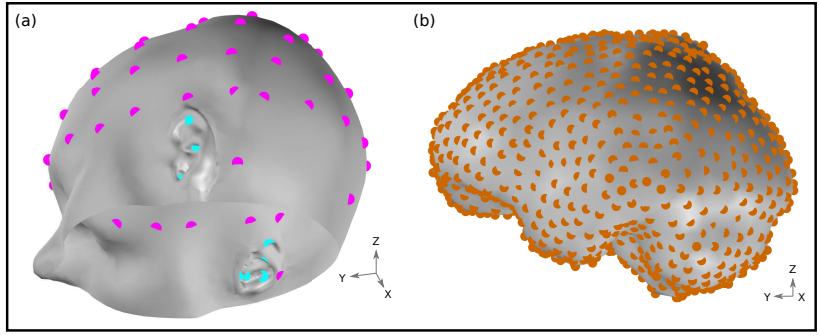}
    \caption{(a) EEG sensor locations on the skin surface of the model. Magenta and cyan markers, respectively, are used to indicate the location of electrodes in a $64$-electrode scalp EEG montage and a left and right ear ear-EEG montage. (b) Neural source space on the course-grained surface of the brain. Source locations are homogeneously distributed with a density of \SI{1.5}{cm^{-2}} and are highlighted in orange.}
    \label{fig sensors sources}
\end{figure*}

\subsection*{Ocular artifact modelling}
Artifacts arising due to blinks and eye movements can be explained in terms of the corneo-retinal dipole (CRD) field. This dipole field arises due to natural charge separation between the cornea and the retina. During eye movements, the CRD rotates around the centre of the eyeball, resulting in a dipole current. During eye blinks, the conductive surface of the inner eyelid sweeps over the cornea, leading to current discharge, which can also be modelled as a current dipole. In \cite{lins1993ocular}, the authors performed dipole fits to electrooculographic data. They found that two-dipole fits (one dipole per eye) explained the data very well (explaining up to around $98\%$ of the total variance). For each type of ocular artifact (vertical saccades, horizontal saccades, blinks), the fitted dipoles shared approximately the same locations. For blinks, the dipoles were approximately aligned in the anterior-posterior direction. For vertical (horizontal) saccades, the dipoles were approximately aligned in the superior-inferior (lateral) direction. We therefore modelled each type of artifact using point current dipoles aligned with the directions reported by Lins, Picton and Scherg \cite{lins1993ocular2}. For convenience, these are displayed in Table \ref{tab:oculardipoles}.

\begin{table*}[]
\centering
\caption{Ocular dipole directions:  Direction of dipoles are given in unit vectors. The z-axis is in approximate alignment with the inferior-superior axis of the head. VS = vertical saccade. HS = horizontal saccade.}
\label{tab:oculardipoles}
\begin{tabular}{@{}llllllllll@{}}
\cmidrule(l){2-10}
          & \multicolumn{9}{c}{Ocular artifcat}                                           \\ \cmidrule(l){2-10} 
          & \multicolumn{3}{c}{Blink} & \multicolumn{3}{c}{VS}  & \multicolumn{3}{c}{HS}  \\ \cmidrule(l){2-10} 
          & x       & y    & z & x   & y   & z & x      & y   & z \\
Left eye  & $-0.304$  & $-0.937$  & $0.174$ & $0.183$  & $-0.159$ & $-0.97$ & $-0.988$ & $-0.156$ & $0.017$ \\
Right eye & $0.432$   & $-0.885$  & $0.174$ & $-0.159$ & $-0.183$ & $-0.97$ & $-0.982$ & $0.156$  & $0.104$ \\ 
\end{tabular}

\end{table*}

In our simulations, we neglected the rider artifact, which is a transient onset effect which occurs at the start of a vertical or horizontal saccade. Similar to blink artifacts, the rider artifact occurs because the eyelid lags behind the motion of the artifact, discharging slightly. In fact, the two blink dipoles can be used to explain the rider artifact.

\section*{Results}
\subsection*{Ear-EEG sensitivity to neural sources}
\subsubsection*{Sensitivity maps}
The source sensor relationship was inspected for the entire surface of the brain through sensitivity maps. Within Matlab ® software, continuous estimates of the sensitivity of EEG over the surface of the brain - sensitivity maps -  were created by projecting the potential difference (sensitivity) arising as a result of each source on to a mesh of the brain, then interpolating sensitivities with the function \texttt{trisurf}, where sensitivities were calculated as the potential difference between two points on the skin surface of the model that correspond to the EEG electrode locations of interest. Usually, ear-EEG electrodes are installed within an earpiece which conforms to the various compartments of the outer ear - the ear canal, concha, and helix. As such, contact points on each of these compartments offer various electrode locations for EEG sensing. Consequently, through the use of a single, simple-to-use device, ear-EEG can be recorded simultaneously from a montage of electrodes. With this in mind, the maximum sensitivity produced by a pair of electrodes from within a montage was used to crease the sensitivity maps. In this way, the full capabilities of a montage could be explored. Sensitivity maps were created for montages of unilateral ear-EEG (single ear), bilateral ear-EEG (dual ear), and scalp EEG (64 channel 10-20 montage). In order to enable clear visualisation of the exponentially dynamic sensitivity map, sensitivities are plot in decibels (dB) relative to an arbitrary value. 

\begin{figure*}
    \centering
    \includegraphics{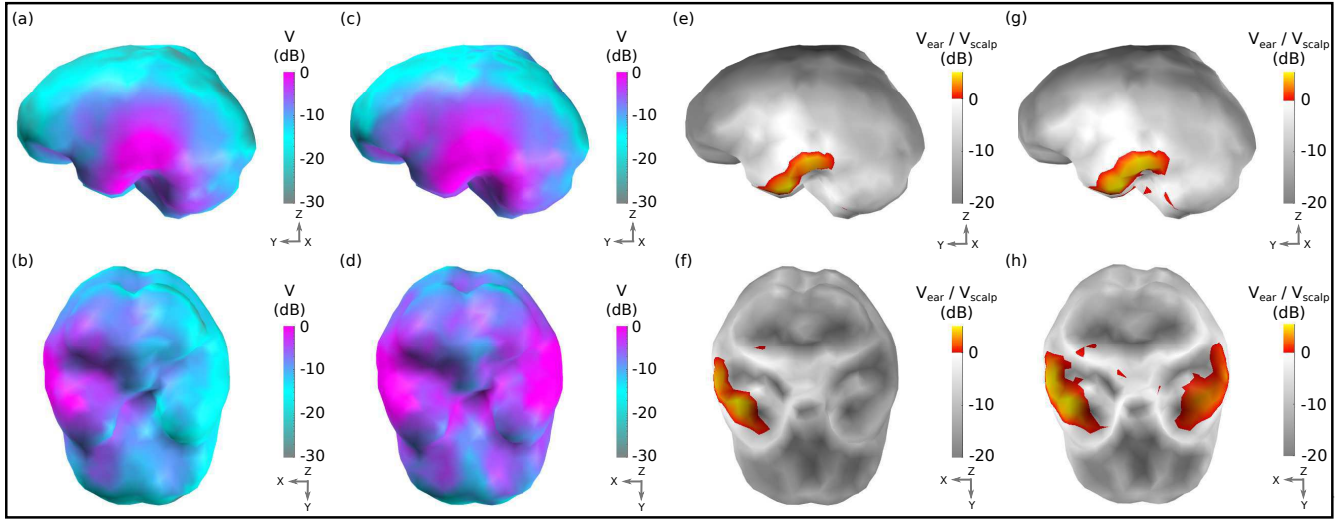}
    \caption{Sensitivity maps for ear-EEG: (a-b) Map for a left ear unilateral ear-EEG montage. (c-d) Map for a bilateral ear-EEG montage. (e-f) Relative sensitivity map for a left ear unilateral montage and a 64-channel scalp EEG montage. (g-h) Relative sensitivity map for a bilateral ear-EEG montage and a 64-channel scalp EEG montage. (a-d) High and low sensitivities, respectively, are displayed in magenta and cyan. (e-h) Significant and moderate signal losses are displayed in grey and white respectively. Signal gains are displayed in red and yellow. (a,c,e,g) Left side of the brain. (b,d,f,h) Inferior surface of the brain.}
    \label{fig topo}
\end{figure*}

Figure \ref{fig topo}a and b displays the sensitivity map for a left ear unilateral ear-EEG montage (displaying the characteristic sensitivity profile for a single earpiece). The highest sensitivities of the montage were exclusively observed in the ipsi-lateral inferior and middle temporal lobe. Decreases in sensitivity were observed for regions surrounding the ipsi-lateral temporal lobe, with the lowest sensitivities observed for sources furthest away from the ipsi-lateral ear; in frontal, central, and posterior, and contra-lateral locations. For the bilateral montage, maximum sensitivities are observed across large portions of the left and right temporal lobe, while the lowest senisvties were observed for frontal, central, and posterior regions close to the mid-line sagittal plane.

In addition to the sensitivity maps described above, relative sensitivity maps were created, where the ear-EEG montage sensitivities were divided by the scalp EEG montage sensitivities (Figure \ref{fig topo}e-h). In this way, the expected signal loss (decrease in signal amplitude) or signal gain (increase) associated with the use of both unilateral and bilateral ear-EEG over scalp EEG could be examined. Meiser \textit{et al.} pioneered this method for CEEG analysis in \cite{Meiser2020}. The relative sensitivity of the unilateral montage is displayed in Figure \ref{fig topo}e-f. On the inferior ipsi-lateral temporal lobe, for a small collection of sources (2\% of the total) there is moderate signal gain. For the majority of the remaining sources, there is a significant signal loss. The median relative sensitivity for all sources for the unilateral ear-EEG montage is \SI{-17}{dB} (25th/75th percentile: \SI{-20}{dB}/\SI{-4}{dB}). For the small portion of sources detected with signal gain, there is a median relative sensitivity value of \SI{2}{dB} (25th/75th percentile: \SI{1}{dB}/\SI{4}{dB}). For the bilateral montage, the regions of significant signal loss are reduced relative to the unilateral montage. The median relative sensitivity was found to be \SI{-10}{dB} (25th/75th percentile: \SI{-15}{dB}/\SI{-4}{dB}). A small portion (5\%) of sources on both temporal lobes were detected with a signal gain, with the median of \SI{2}{dB} (25th/75th percentile: \SI{1}{dB}/\SI{3}{dB}).

\subsubsection*{Channel sensitivity analysis}
\begin{figure*}[t]
    \centering
    \includegraphics{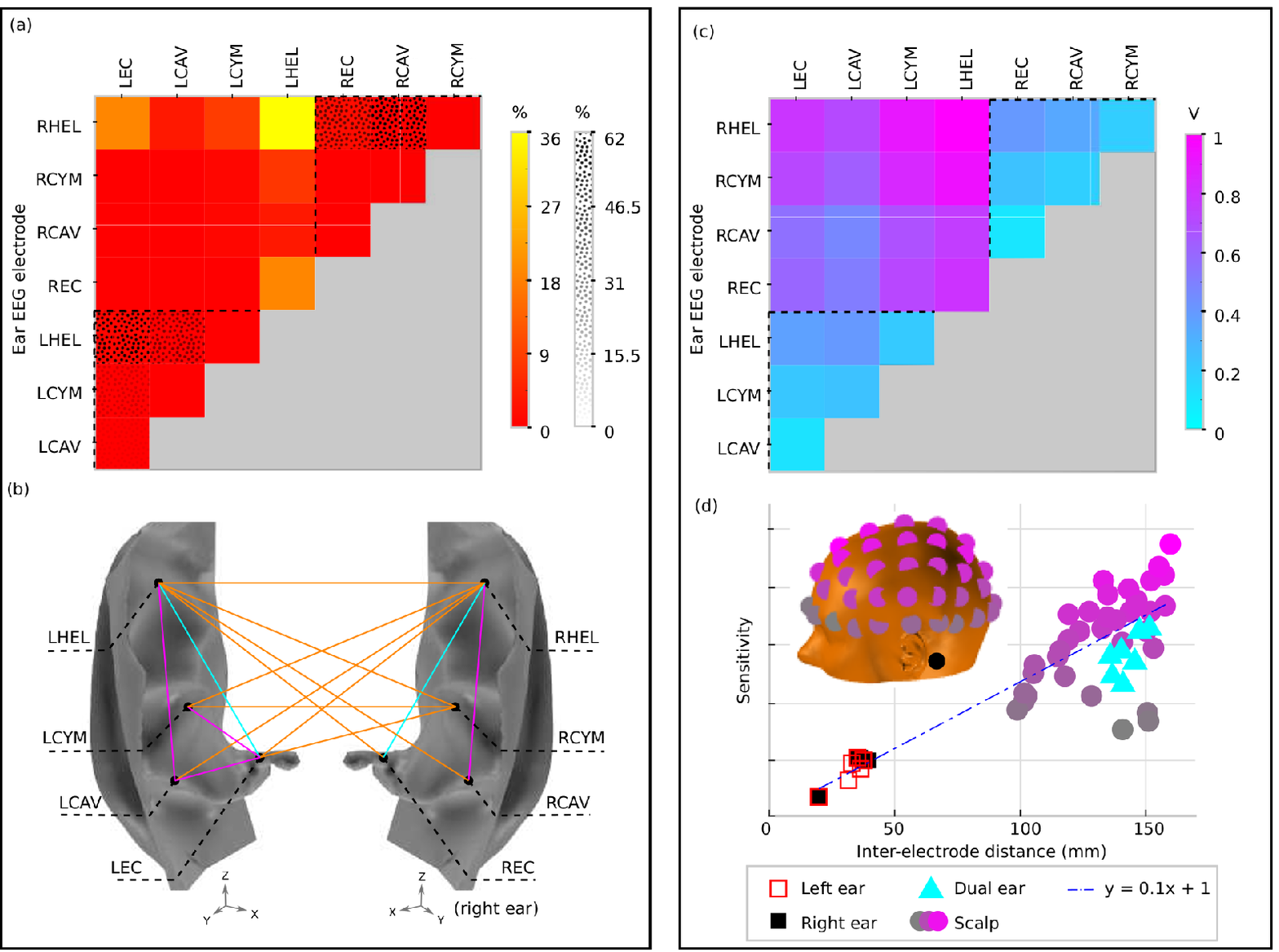}
    \caption{Channel prevalence and average sensitivity analysis for unilateral and bilateral montages of ear-EEG. (a) Channel prevalence for unilateral and bilateral montages of ear-EEG. Bilateral prevalence is shown in a colour scale, while unilateral prevalence for the left and right ears is shown a in texture scale. Dashed lines separate single ear and bi-ear values on the heat-map. (b) Channel prevalence for unilateral and bilateral ear-EEG montages visualised on the surface of the ear. For channels that recorded the highest sensitivity within a montage for at least one source ($>$ 0\%), a coloured line connects the channel's electrodes. Magenta, orange, and cyan, respectively, are used to represent channels which were prevalent exclusively within a unilateral montage, exclusively within a bilateral montage, and both unilateral and bilateral montages. (c) Normalised average sensitivity for unilateral and bilateral montages. Dashed lines separate single ear and bi-ear values on the heat-map. (d) Relationship between sensitivity (normalised) and inter-electrode distance for both ear-EEG (unilateral and bilateral) and scalp EEG. A linear fit of the data is shown in a blue dashed line. Scalp EEG channel locations are indicated in the inset and coloured corresponding to their sensitivity values. All channels displayed are referenced to the average of the sensitivities of the mastoid electrodes; the location of the left mastoid electrode is indicated by a black circle on the inset, while the right mastoid electrode is hidden on the equivalent position on the right side of the head.}
    \label{fig bi_uni}
\end{figure*}

While ear-EEG can conveniently be recorded simultaneously from a montage of electrodes spread across the various outer-ear positions, the characteristic performance of individual channels within the montage is also of interest since, in certain scenarios, it could be desirable to minimise the number of ear-EEG electrodes, e.g., when space for other sensors might be required. The various channels available to ear-EEG have been examined separately in the following analysis. Each individual channel's prevalence has been measured, where each channel's prevalence is equal to the percentage of brain sources for which the given channel recorded the highest sensitivity. The prevalence measure indicates the relative utility of the channel within its montage. Only sources for which at least one channel recorded a sensitivity exceeding a threshold were included in the analysis. The threshold was set to \SI{10}{dB} below the peak amplitude of the 64 channel scalp channels, where the peak amplitude for each channel is the highest amplitude recorded from all 990 brain surface sources. As such, only sources which were favourably detected by ear-EEG were included in the analysis. Figure \ref{fig bi_uni}a displays the channel prevalence in the form of a heat map for both unilateral and bilateral montages (indicated by a dotted pattern fill and colour shading, respectively). The analysis of channel prevalence was introduced for CEEG in \cite{Meiser2020}. Within the left unilateral montage (bottom left of the heat map, separated by a dashed line) the helix to ear canal channel recorded the maximum sensitivity for 56\% of selected sources. Helix to cavum, and cymba to ear canal were the next most prevalent (30\% and 10\%, respectively), followed by cavum to ear canal (3\%). Helix to cymba and cymba to cavum failed to record the highest sensitivity for a source. Within the right ear unilateral montage (top right of the heat map, separated by a dashed line) the helix to cavum was the most prevalent (62\%), with the helix to ear canal recording the highest sensitivity for the remaining sources (38\%). Within the bilateral montage, which included all available electrodes on both the left and right ears, the left helix to right helix channel recorded the most maximum sensitivities (36\%) for a single channel, while the bi-ear helix to ear canal channels (left helix - right ear canal / right helix to left ear canal) recorded the maximum sensitivity for 20\% of sources each (Figure \ref{fig bi_uni}a). The left helix to right cymba and right helix to left cymba were the next most prevalent, recording 10\% and 9\% of maximum sensitivities, respectively. Figure \ref{fig bi_uni}b displays the complete set of channels which recorded the maximum sensitivity at least once. To supplement the channel prevalence analysis, the average sensitivity for each ear-EEG channel was also calculated. As with the prevalence analysis, average sensitivities were based only on sources which satisfied the threshold condition described above. The helix to helix channel recorded the highest average signal amplitude, however the majority of bi-ear channels recorded similar average amplitudes. The single ear channels exhibited lower average signal amplitude; the lowest average amplitude was recorded by the left and right single ear cavum to ear canal channels. 

In order to further characterise ear-EEG, for a selection of ear- and scalp-EEG channels, channel sensitivities were plotted against respective inter-electrode distance (Figure \ref{fig bi_uni}d). In this analysis, in order to enable comparison between scalp EEG and ear-EEG, the sensitivity of each channel was indicated by the number of sources for which the channel sensitivity exceeded the previously described threshold. For the scalp EEG channels, a linked mastoid referencing system was used, while for ear-EEG, left and right ear channels were referenced to the ipsi-lateral helix, and bi-ear channels to the contra-lateral helix. The inter-electrode distance for the linked mastoid referenced scalp EEG channels was calculated as the average of the distance of the primary electrode from both mastoid electrodes. Both the sensitivity and inter-electrode distance of single ear channels are lower than those of the bi-ear and scalp EEG channels, leading to the ratio of mean channel sensitivity for left ear, right ear, and bi-ear-EEG relative to the mean channel sensitivity of scalp EEG, respectively, of 0.3, 0.3, and 0.9. A linear trend with equation of best fit,  $y = 0.1x  + 1$, where $y=$ sensitivity and $x=$ inter-electrode distance, was found for the data displayed in the plot. 

\subsection*{Artifact modelling}
\subsubsection*{Ocular artifacts}
\begin{figure*}[h!]
    \centering
    \includegraphics{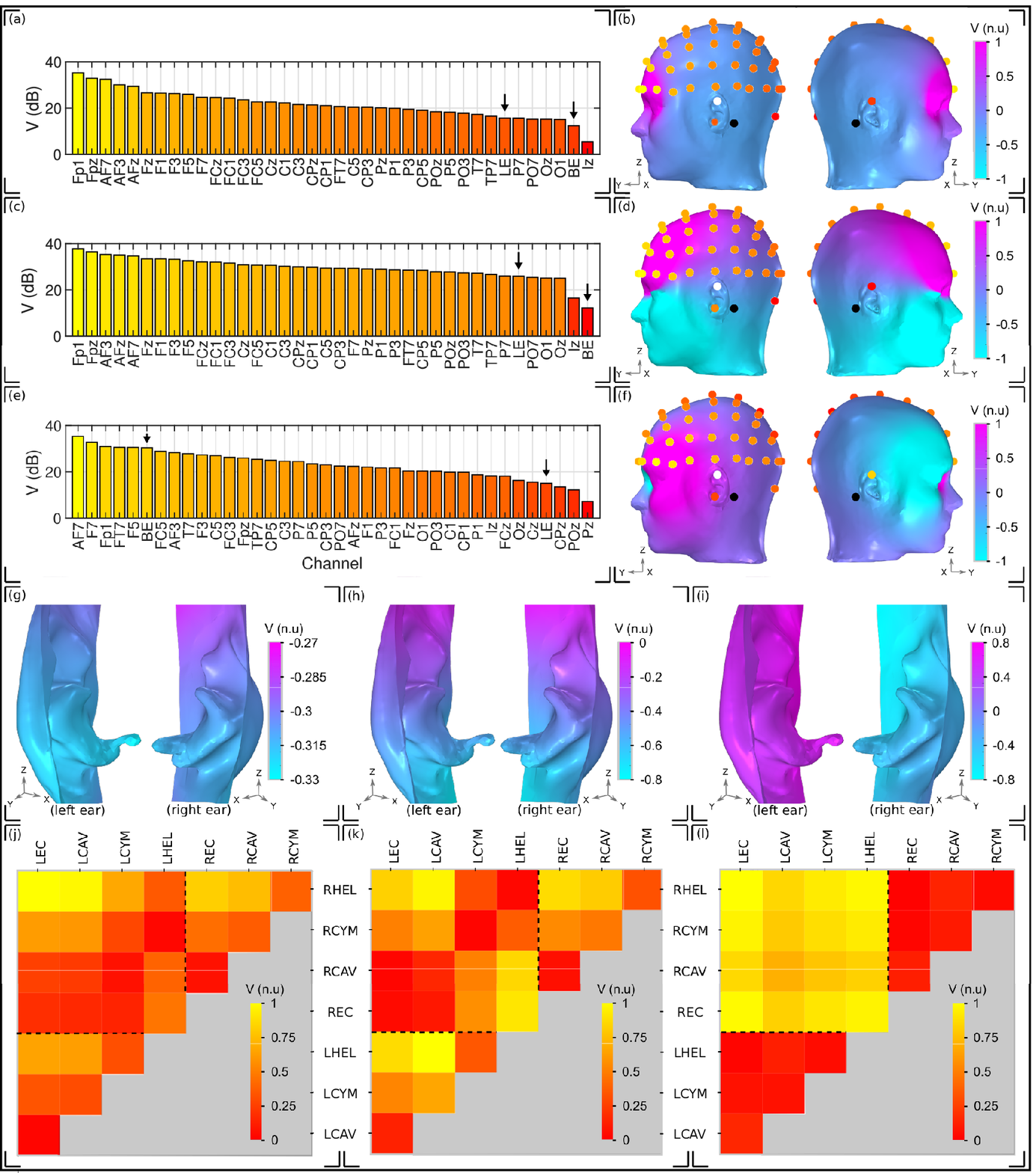}
    \caption{Ocular artifacts. (a,c,e) Sensitivities of the left hemisphere scalp EEG channels in addition to a unilateral left ear (LE) and bilateral (BE) ear-EEG channel, for (a) blinks, (c) vertical saccades, and (e) horizontal saccades. (b,d,f) Head surface potential topography and EEG sensitivities arising due to (b) blinks, (d) vertical saccades, and (f) horizontal saccades. Scales are normalised uniformly across each of the head surface topographies such that an inter-artifact comparison is possible. White (black) circles indicate ear (scalp) reference electrodes. (g,h,i) Left and right ear surface potential topography arising due to (g) blinks, (h) vertical saccades, and (i) horizontal saccades. Potential scales are shared with the scalp topographies, however the colour scale has been changed to enable clear visualisation of the potential topography over each ear surface. (j,k,l) Sensitivity of ear-EEG channels arising due to (j) blinks, (k) vertical saccades, and (l) horizontal saccades. The potential scale has been normalised for each individual plot; potential scales are not shared between artifacts.}
    \label{fig ocular artifacts}
\end{figure*}

 The sensitivity of different configurations of ear-EEG in the presence of three common types of ocular artfiact, blinking, vertical saccade, and horizontal saccade, was investigated. The sensitivity of ear-EEG to ocular artfictas were bench-marked against that of scalp EEG in Figure \ref{fig ocular artifacts}a-c. Specifically, sensitivities for two ear-EEG channels; single-ear left helix to left ear canal (LEH-LEC) and bi-ear left helix to right helix (LHEL-RHEL) are displayed alongside sensitivities of scalp EEG channels. Since the scalp EEG channel sensitivities are symettric about the midline sagittal plane, the inclusion of only a single hemisphere EEG channels was sufficient to capture the variations in scalp EEG sensitivities to ocular artifacts. As a result of the large range (multiple orders of magnitude) of channel sensitivities, values in dB relative to an arbitrary value reference were displayed. For all three artifacts, sensitivity values are reported with respect to the same reference value to enable between-artifact comparison.

For the blink artifact (Figure \ref{fig ocular artifacts}a), the maximum scalp EEG sensitivity was recorded by the FP1 channel and the lowest by the Iz channel; at a value \SI{33}{dB} below that of FP1. Regarding ear-EEG, the blink artifact resulted in a larger potential difference in the LHEL-LEC channel relative to the LHEL-RHEL channel, with a difference between the sensitivities of \SI{3}{dB}. Relative to scalp EEG, the LHEL-LEC channel was most similar ($<$ \SI{1}{dB} difference) to scalp channels with a lateral positioning (P7, TP7), while the LHEL-RHEL channel was most similar ($<$ \SI{1}{dB} difference) to the posterior scalp channel O1. The LHEL-LEC and LHEL-RHEL sensitivities were amongst the least sensitive out of the selection of scalp and ear-EEG channels analysed (7th and 2nd least sensitive out of 37 channels, respectively). The resultant potential topography on the surface of the head with overlaid EEG channel topography is shown in Figure \ref{fig ocular artifacts}b. For the vertical saccade artifact, the maximum scalp EEG sensitivity was recorded by the FP1 channel and the lowest by the Iz channel; at a value \SI{26}{dB} below that of FP1. The LHEL-LEC channel was most similar to lateral and posterior scalp EEG channels, P7 and POz ($<$ \SI{26}{dB} difference), while the LHEL-RHEL channel was most similar to the posterior inferior scalp EEG channel, Iz ($<$ \SI{5}{dB} difference). The LHEL-LEC channel was \SI{14}{dB} more sensitive relative to the LHEL-RHEL channel. The LHEL-LEC and LHEL-RHEL sensitivities were amongst the least sensitive out of the selection of scalp and ear-EEG channels analysed (6th and 1st least sensitive, respectively). The topography of the vertical saccade potential over the surface of the scalp is shown in Figure \ref{fig ocular artifacts}d. For the horizontal saccade artifact, the maximum scalp EEG sensitivity was recorded by the AF7 channel and the lowest by the central and parietal scalp EEG channels along the midline saggital plane, Cz and CPz ($<$ \SI{2}{dB} difference), while the LHEL-RHEL channel was most similar to frontal-lateral and central-lateral channels, F5 and FC5 ($<$ \SI{2}{dB} difference). The LHEL-LEC channel was \SI{15}{dB} less sensitive relative to the LHEL-RHEL channel. The LHEL-LEC sensitivity was amongst the least sensitive out of the selection of scalp and ear-EEG channels analysed, while the LHEL-RHEL sensitivity was amongst the most sensitive (4th least sensitive and 6th most sensitive, respectively). The topography of the horizontal saccade potential over the surface of the scalp is shown in Figure \ref{fig ocular artifacts}d. The mean, maximum, and minimum sensitivity for the blink, vertical saccade (VS), and horizontal saccade (HS) artifact were also calculated for each montage; within the linked mastoid scalp EEG channels, sensitivities were, blink: \SI{26}{dB}, \SI{35}{dB}, \SI{5}{dB}, VS: \SI{31}{dB}, \SI{38}{dB}, \SI{17}{dB}, and HS: \SI{27}{dB}, \SI{35}{dB}, \SI{7}{dB}. For the single ear channels, the sensitivities were, blink: \SI{13}{dB}, \SI{16}{dB}, \SI{1}{dB}, VS: \SI{24}{dB}24, \SI{27}{dB}, \SI{15}{dB}), and HS: \SI{20}{dB}, \SI{22}{dB}, \SI{14}{dB}. For the bi-ear channels, the sensitivities were, blink: \SI{20}{dB}, \SI{22}{dB}, \SI{14}{dB}, VS: \SI{20}{dB}, \SI{22}{dB}, \SI{14}{dB}, and HS: \SI{20}{dB}, \SI{22}{dB}, \SI{14}{dB}. The surface potential on the left and right ears for each artifact are displayed in Figure \ref{fig ocular artifacts}h-j. For each artifact, Figure \ref{fig ocular artifacts}k-m display the normalised sensitivities for the various ear-EEG channels. 

\section*{Discussion}

\subsection*{Sensitivity to neural sources}

The sensitivity of both unilateral (single ear) and bilateral (bi-ear) montages to neural sources across the entire brain surface was examined. While unilateral measurements are confined to measuring potential differences over the small region of the ear, bilateral montages enable measurement between the left and right ears. In this way, the bilateral montage increases the inter-electrode distance, and therefore the potential difference,  as a result of the physical laws governing EEG. The simulations reported enable examination of the effect of larger inter electrode distance associated with bilateral montages on the ear-EEG sensitivity profile over the brain surface. The results reveal that, although the bilateral montage increases sensitivity, large scale variations in sensitivity are similar to those found for the unilateral montage in the primary region of sensitivity - the temporal lobe - and the regions of lowest sensitivity in proximity to the mid-line sagittal plane. The regions where the benefits of larger inter-electrode distance were observed were those adjacent to the temporal lobe. Such differences between the unilateral and bilateral montages were observed for both the regular sensitivity maps (ear-EEG sensitivity) and the relative sensitivity maps (ear-EEG sensitivity bench marked against scalp-EEG sensitivity).

Since EEG is conventionality recorded with the scalp-EEG setup, it useful to compare the amplitude of ear-EEG signal to that of scalp EEG. Therefore, relative sensitivity maps were also created. Ear-EEG resulted in an increase in signal amplitude in small regions in the temporal lobe, while adjacent regions mostly exhibited moderate decrease in signal amplitude. In regions furthest away from the ear-EEG electrodes, the ear-EEG amplitude is likely to be significantly smaller than that of scalp EEG. The results for sources in and adjacent to the temporal lobe suggests that ear-EEG can be expected to record EEG amplitudes similar to those seen in scalp EEG. Since temporal lobe neural activity is known to correspond to important auditory and visuo-auditory processing, amongst other functionality, the use of ear-EEG in applications such as enhanced, 'smart' hearing aids is strongly supported by these results. 

Analysis of the various channels within a unilateral montage showed that the channels maximising the space available to the montage, while also utilising the helix electrode (proximity to the brain), recorded the highest amplitude for the majority of sources. These were, the helix to ipsi-lateral ear canal and helix to ipsi-lateral cavum channels. For the bilateral montage, the same trend was observed, with the helix to contra-lateral ear canal channels and helix to contra-lateral helix channels producing the highest sensitivities. As such, a reduction of the ear-EEG montage size could see the use of the helix and ear canal electrodes prioritised. Despite the dominance of the helix and ear canal electrode, in both the unilateral and bilateral montages, multiple of the electrode locations tested contributed to the highest sensitivity for at least one source; this indicates that a multi-electrode array of electrodes within the ear-EEG montage could be beneficial. 

For additional insight, the average sensitivity of the channels was also calculated. Calculations revealed that, although the channels that produced the highest prevalence also recorded the highest average sensitivity, the sensitivities were comparable for channels comprising solely ipsi-lateral pairings, or solely contra-lateral pairings. The similarity in sensitivity between channels suggests that the positioning of the electrodes can be flexible. Such flexibility may be leveraged in Hearable devices, which employ a selection of health monitoring sensors, including EEG electrodes, during the optimisation of positioning of sensors with varying modalities and positional requirements \cite{goverdovsky2017hearables}. The expected signal amplitude of various ear-EEG and scalp EEG channels were then examined in relation to their inter-electrode distance. In general, for both scalp and ear-EEG, the expected signal amplitude was linearly proportional to inter-electrode distance. The revealed relationship between expected signal amplitude and inter-electrode distance could be used during the design process for various wearable/Hearable devices which utilise the ear and scalp surfaces. 

\subsection*{Sensitivity to ocular artifacts}
For prospective general purpose EEGs, the sensitivity to ocular artifacts is an important design factor as a result of the regularity of eye movements. The sensitivity to ocular artifacts for ear and scalp EEG was demonstrated through both single channel sensitivity measurements and topographical plots of potential over the surface of the head. ear-EEG sensitivities were calculated for a characteristic channel from both a unilateral montage and a bilateral montage. The sensitivity to ocular artifacts for the single ear channel relative to scalp EEG channels was observed to be generally low, evidenced through channel sensitivities of ear-EEG matching those of scalp EEG channels which are amongst the least severely effected by ocular artficats. There was further evidence of low ear-EEG sensitivity in the topographical plots, where, relative to the scalp surface, the ear surfaces exhibited low changes in potential, resulting in low limits of potential difference within the unilateral ear-EEG montage. The bi-ear channel was also low in sensitivity for the blink and vertical saccade, once again evidenced through similarity to low sensitivity scalp EEG channels, and minimally varying topography between the two ears. An exception occurred for the bilateral channel and the horizontal saccade, which was equally large in amplitude in the bilateral ear-EEG channel and the worst affected frontal-lateral scalp EEG channels. The topography for the horizontal artifact reveals a large potential difference between the two ears. 

The variation in ocular artifact amplitude between various ear-EEG channels was also examined. The maps show expected trends, where channels with larger inter-electrode distance have generally larger amplitudes, however, there also variations in sensitivity which arise due to variations in channel orientation relative to the dipole field for the artifact, e.g., the \SI{14}{dB} increase in sensitivity for the LHEL-LEC channel relative to the LHEL-RHEL channel, despite a much smaller inter-electrode distance. Such characteristic variations within the ear-EEG montages could be used in the detection of artifact within the ear-EEG array; this highlights another benefit of utilising the multi-electrode ear-EEG array, as opposed to single channels. 
\subsection*{Suggestions for ear-EEG}
The results within this study have, for the first time, mapped the sensitivity of various configurations of ear-EEG to both neural and ocular sources in a detailed manner. Such source-sensor mapping, while serving to provide novel insight into the ear-EEG sensitivity profile, could also be used to support experimental ear-EEG, either in existing literature, or in prospective studies, with regard to both EEG detection and sensitivity to ocular artifacts. Additionally, the methods employed within this study can be adopted with reasonable ease by other researchers for the purpose of conducting new ear-EEG simulations, possibly for cases where ear-EEG / source configurations that have not been simulated within this study are of interest. 

With regard to suggestions for the application of ear-EEG provided by this study, simulations of neural source sensitivity have conclusively supported the use of ear-EEG for the detection of neural activity originating from within the temporal lobe. In both the unilateral and bilateral cases, the ear-EEG recorded higher or similar amplitudes to conventional scalp EEG within these regions. This suggests that existing protocols for EEG detection could be used with ear-EEG, without the need for significant changes to the protocol, and with equally likely success. In fact, as a result of the wearability of ear-EEG, existing auditory EEG protocols could feasibly be enhanced to include more novel real world recording scenarios, as demonstrated in a 'smart helmet with ASSR' study \cite{von2016smart}. Since moderate decreases in amplitude were also observed for ear-EEG in brain regions adjacent to the temporal lobe, there is also support for the use of ear-EEG for a wider range of applications which require monitoring of neural activity originating within these regions. However, in such applications, it is more likely that the bilateral montage would be more suitable, since the amplitude of the EEG signal would be greater.

Despite the low amplitude of signal detected from areas furthest away from the temporal lobes, e.g., the posterior regions of the brain, there is published experimental support for ear-EEG detection of neural activity from such regions, such as the visual cortex in the occipital lobe (e.g., \cite{goverdovsky2017hearables,kidmose2013study}). Within the experimental studies, the successful detection of visual ERPs through ear-EEG have been possible despite smaller amplitudes (as predicted within the simulations within this paper). The reason for this is the lower noise amplitudes within the ear-EEG. An example of such balancing in signal and noise amplitude reduction in ear-EEG is provided through the simulations of both neural and ocular source sensitivities in Figure \ref{fig ocular artifacts} in this paper. Such results support the continued use of ear-EEG in the detection of neural activity from regions further away from the ear, despite lower expected signal amplitude. 

\subsection*{Limitations and future work}

The first limitation of the presented study is the absence of absolute sensitivity predictions, i.e., predictions of the amplitude of ear-EEG that could be seen in recordings. Absolute predictions could be made possible through calibration of the dipole amplitudes, however, the approach adopted within this study focused on differences in potential between relatively large distances over the brain surface. In this way, meaningful comparisons between more general variations in source-sensor relationship were provided.

A second limitation concerns minor misrepresentations in the dielectric distribution within the presented volume conductor model. More specifically, the absence of differentiation between the grey and white matter, and muscle conductivity anisotropies. The effects of these minor misrepresentations were made less severe through a focus on general variations in source sensor relationship \cite{vorwerk2014guideline}. A more precise extension of the presented model could be feasibly created through the utilisation of separate meshes for the grey and white matter, as provided within the MIDA data set \cite{iacono2015mida}, and the integration of muscle anisotropy data into the current model \cite{aaron1997anisotropy}. However, such modelling would require additional computational resources, while its predictions would yield limited gains in accuracy as a result of the lack of subject-specific anatomical detail. 

When it comes to providing a detailed analysis of the source-sensor relationship, an analysis of the sensitivity to source orientation, position, and distance has not been systematically provided. Meiser \textit{et al.} \cite{Meiser2020} highlighted the utility of such an analysis in the context of CEEG. The fine grained approach to modelling employed within this paper (mapping the sensitivity to individual dipoles on the brain surface) did reveal the sensitivity of ear-EEG to small variations in dipole orientation, position, and distance however, as evidenced by the large variations in sensitivity for adjacent sources (see the isolated peak in the relative sensitivity profile for small regions on the inferior surface of the brain for both the unilateral and bilateral ear-EEG montages). Such variations point to the large variations in sensitivity which stem from the interplay between source orientation, position, and distance. A systematic analysis of such sensitivities would be a logical extension of the presented study.  

A second extension of the presented study would be the simulation of sensitivity for an ear-EEG array of a higher density. While the current study analysed four ear-EEG electrode locations (ear canal, cavum, cymba, and helix), an analysis of a higher population of the outer ear surface could enable a meaningful analysis of the effect of source orientation, position, and distance on the ear-EEG sensitivity, since a higher density array would enable more variations in channel geometry. In turn, the analysis of a high density array would also be useful in the context of predicting the ability of ear-EEG enabled source separation/localisation.  

\section*{Conclusion}
The source-sensor relationship in ear-EEG, for both neural and ocular artifact sources, has been analysed for the first time. The results have provided conclusive evidence for the use of ear-EEG in applications concerning the monitoring of neural activity originating from within the temporal lobes. Further, evidence has been provided for equal SNR between ear-EEG and scalp EEG in the presence of ocular artifacts. The reported results can also be used as a reference for relative ear-EEG amplitudes for various neural and ocular sources, supporting both existing and prospective experimental ear-EEG studies. Future work will exploit the utility of physics modelling to provide further insight into the sensitivity of ear-EEG to both neural and artifact sources. 

\section*{Acknowledgements}

Metin Yarici was supported by the Racing Foundation grant 285/2018 and the MURI/EPSRC grant EP/P008461. Michael Thornton was supported by the UKRI CDT in AI for Healthcare \href{http://ai4health.io}{http://ai4health.io} (Grant No. P/S023283/1).

\section*{Appendix}
\subsection*{Volume conductor model processing}
Forward modelling within COMSOL is based on the application of physics equations to geometrical objects. As such, geometrical artifacts such as non-manifold vertices and self-intersecting edges and surfaces are incompatible with forward modelling within COMSOL. The high spatial resolution of the MIDA model includes vast amounts of the aforementioned geometrical artifacts, in addition to sub-millimetre surface detail which is not required for the purpose of EEG forward modelling. In order to remove both artifacts and unnecessary surface detail, data were imported into Autodesk Meshmixer and edited. Editing was conducted via a mixture of mesh operations including smoothing, remeshing, reduction, removal, filling, Boolean joining, and grouping. The final edited geometries are smooth, shell-like structures which describe the outermost surfaces of a tissue domain. In order to build each of the final geometries, several of the segmented tissues from the MIDA model were used (see Table \ref{tab:tissue_mida}). 
Once the editing was complete, the geometries were subsequently imported into COMSOL. Utilising a 'finer' mesh size in COMSOL, the final volume conductor model, including the enclosing sphere, consisted of a total of $80874$ face elements, and $63046$ vertex elements.

\bibliographystyle{IEEEtran}
\bibliography{references}
\begin{table*}
\centering
\caption{MIDA files used in the construction of the presented volume conductor model.}
\label{tab:tissue_mida}
\resizebox{\textwidth}{!}{%
\begin{tabular}{@{}lclllll@{}}
\cmidrule(l){2-7}
 &
  \multicolumn{6}{c}{Tissue} \\ \cmidrule(l){2-7} 
 &
  Bone &
  \multicolumn{1}{c}{Brain} &
  \multicolumn{1}{c}{Skin} &
  \multicolumn{1}{c}{Parotid Glands} &
  \multicolumn{1}{c}{Eyes (vitreous humour)} &
  \multicolumn{1}{c}{Air} \\
MIDA files &
  \multicolumn{1}{l}{\begin{tabular}[c]{@{}l@{}}Skull diploe,\\ Skull inner table,\\ Skull outer table,\\ Skull, \\ Teeth,\\ Spinal cord, \\ C1, \\ C2, \\ C3\end{tabular}} &
  \begin{tabular}[c]{@{}l@{}}Gray matter,\\ White matter,\\ Cerebellum gray matter,\\ Cerebellum white matter, \\ Brainstem midbrain,\\ Brainstem pons\end{tabular} &
  \begin{tabular}[c]{@{}l@{}}Adipose tissue,\\ Subcutaneous adipose tissue,\\ Auditory canal,\\ Auricular cartilage,\\ Epidermis and dermis\end{tabular} &
  Parotid gland &
  Eye vitreous &
  \begin{tabular}[c]{@{}l@{}}Mastoid,\\ Nasal pharynx,\\ Oral cavity\end{tabular}
\end{tabular}%
}
\end{table*}
\end{document}